# Theoretical Study of Thermopower Behavior of LaFeO$_3$ Compound in High Temperature Region

Saurabh Singh,[1, a)] Shivprasad S. Shastri[1] and Sudhir K. Pandey[1]

[1]*School of Engineering, Indian Institute of Technology Mandi, Kamand 175005, Himachal Pradesh, India.*

a)Corresponding author: saurabhsingh950@gmail.com

**Abstract.** The electronic structure and thermopower ($\alpha$) behavior of LaFeO$_3$ compound were investigated by combining the *ab-initio* electronic structures and Boltzmann transport calculations. LSDA plus Hubbard U (U = 5 eV) calculation on *G-type* anti-ferromagnetic (AFM) configuration gives an energy gap of ~2 eV, which is very close to the experimentally reported energy gap. The calculated values of effective mass of holes ($m^*_h$) in valance band (VB) are found ~4 times that of the effective mass of electrons ($m^*_e$) in conduction band (CB). The large effective masses of holes are responsible for the large and positive thermopower exhibited by this compound. The calculated values of $\alpha$ using BoltzTraP code are found to be large and positive in the 300-1200 K temperature range, which is in agreement with the experimentally reported data.

## INTRODUCTION

In the past several decades, LaFeO$_3$ (LFO) compound has been studied extensively due to its interesting physical properties such as magnetic, optical, electrical, and thermoelectric properties. This material can be utilized for the various applications like solid oxide fuel cells, sensors, magnetic storage etc.[1] Among various physical properties exhibited by this compound, thermoelectric properties of this system has not been explored much to the best of our knowledge. The ground state structure of this system is reported as *G-type* AFM insulator.[2,3] The experimental energy gap, E$_g$, reported from the optical measurement is ~2.1 eV.[4] LFO compound has the orthorhombic crystal structure described by the *Pbnm* space group, and shows the highest Neel temperature (T$_N$ ~740 K) among all orthoferrites.

For this compound, Minh *et. al* have experimentally reported the large and positive value of $\alpha$, ~360 µV/K at ~520 K, and the values remains large and positive up to high temperature i.e. ~165 µV/K at ~1173 K.[5] The observation of large $\alpha$ values suggests that this system can be useful for the thermoelectric application in high temperature region. To the best of our knowledge, the detailed understanding of experimental thermopower data is still lacking in the literature. In order to understand the mechanism which is responsible for showing the large and positive $\alpha$ behavior by this compound, a detailed theoretical investigation is required. This gives motivation to explore the $\alpha$ behavior of this compound using theoretical tools.

In the present work, electronic structure and thermopower behavior of LFO compound in the high temperature region have been studied using the first principle density functional theory and BoltzTraP code. We have calculated the total density of states (TDOS) corresponding to the *G-type* AFM unit cell. The effective masses of holes and electrons in the VB and CB have been estimated for the qualitative understanding of the large and positive $\alpha$ exhibited by this compound. Using the BoltzTraP code, the values of $\alpha$ is also calculated in the 300-1200 K temperature range.

## COMPUTATIONAL DETAILS

Electronic structure calculations were performed by using the full-potential linearised augmented plane-wave (FP-LAPW) method within the density functional theory (DFT) implemented in WIEN2k code.[6] For the calculation of $\alpha$, BoltzTraP code was used.[7] The exchange correlation function within local density approximation (LSDA) was used and on site Coulomb interaction strength, $U = 5$ eV, among the Fe $3d$ electrons were taken into account so that it gives the energy gap of ~ 2 eV, which is very close to the experimentally reported $E_g$.[8] The experimental lattice parameters ($a = 5.570$ $A^0$, $b = 5.532$ $A^0$, and $c = 7.890$ $A^0$) of orthorhombic crystal structure described by *Pbnm* space group (No. 62), were used for the calculations.[5] The muffin-tin radii were fixed to 2.49, 1.95 and 1.68 Bohr for the La, Fe and O atoms, respectively. For the calculations of electronic and transport properties, the values of k-point mesh were set to the size of 30*30*30. The self-consistency was achieved by demanding the convergence of the total energy/cell and charge/cell to be less than $10^{-6}$ Ry and $10^{-4}$ electronic charge, respectively.

## RESULTS AND DISCUSSIONS

To understand the electronic structure and thermopower behavior of this compound, we have performed the self-consistency field calculations for both ferromagnetic (FM) and *G-type* AFM structure. For *G-type* AFM solution, the value of total converged energy was found to be ~245 meV f.u.$^{-1}$ lower than that of FM solution. This suggests that ground state structure of this compound is *G-type* AFM, which is also in accordance with the literature. Therefore, we have further calculated the TDOS for the G-type AFM structure.

Figure 1a shows the TDOS plot obtained corresponding to the *G-type* AFM phase for the LFO compound. From the TDOS plot, it is clearly observed that LFO compound is an insulating material with $E_g$ value equal to ~2 eV, which is nearly same as experimentally reported $E_g$.[4] As the temperature of the system increases, the fraction of electrons, given by $e^{-E_g/K_B T}$, gets thermally excited from the top of VB and reach to the bottom of CB. These thermally excited electrons from VB to CB, mainly contribute to the electronic transportation.[9]

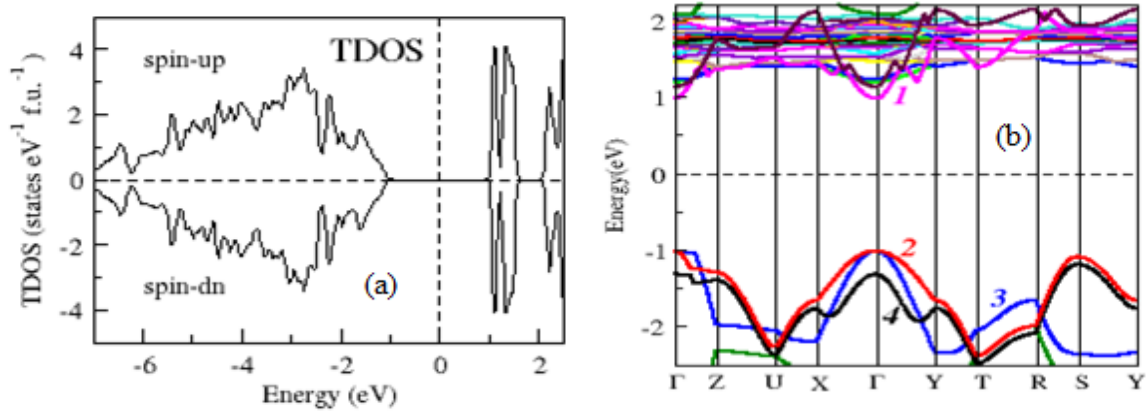

FIGURE 1. (a) Total density of states and (b) Electronic band structure along the high symmetric points ($\Gamma$, Z, U, X, Y, T, R, and S) of LaFeO$_3$ compound.

For a given material, the magnitude of $\alpha$ depend on the effective masses of charge carriers, carrier concentration and the type of charge carriers, whereas the sign of $\alpha$ which gives information about the type of materials (*p-type* or *n-type*) are decided by the effective masses of charge carriers. Thus, the estimation of effective masses of charge carriers plays an important role in the qualitative understanding of the $\alpha$ behavior of the materials.[9,10] The values of effective masses of charge carriers can be calculated from the electronic band structure. Therefore, we have also calculated the dispersion curves for different *k*-values along the high symmetric $\Gamma$, Z, U, X, Y, T, R, and S points. It is clearly observed from the dispersion curves, shown in Fig. 1b, that the top of VB and bottom of the CB lie at the same $\Gamma$ point, which suggest that system is direct band gap insulator with energy gap of ~2 eV. The nature of direct band gap for this system obtained from our calculation is also consistent with the literature.[4] Top of the VB at

Γ point are doubly degenerate (Band *2* and *3*) and this degeneracy are completely lifted on moving from Γ to X and Y directions. The band *1* in the CB is non-degenerate at Γ point.

At finite temperature, the fraction of electrons get thermally excited from the doubly degenerate bands *2* & *3* (at Γ point of the VB) to the non-degenerate band *1* (at Γ point of the CB) and participate in the transport properties. The calculated values of $m^*_h$ and $m^*_e$ along the different high symmetric points are shown in Table 1. The notation in Table 1, Γ- Γ **Y** represents the calculated values of $m^*_h$ and $m^*_e$ at Γ point along the Γ to Y direction, and similarly for others. The values of $m^*_h$ in VB (band *2* & *3*) are found to be larger than the $m^*_e$ in CB (band *1*) for motion of charge carriers along all high symmetric points. The values of $m^*_h$ in the VB for the band *2* (along Y, X, T and Z points) are almost 4 times, whereas $m^*_h$ for the band *3* (along Y, X, T and R points) are found to be ~3 times of the $m^*_e$ in band *1* of the CB. The large values of $m^*_h$ in VB suggests that this compound exhibits large and positive values of α.

**TABLE 1**. The effective mass of holes and electrons at Γ Point and along different high symmetric points.

| High-Symmetric point | Effective Mass (m*/me) | | CB |
|---|---|---|---|
| | VB | | |
| | Band *2* | Band *3* | Band *1* |
| Γ- Γ **Y** | 3.96 | 3.28 | 0.94 |
| Γ- Γ **X** | 3.94 | 3.24 | 0.95 |
| Γ- Γ **T** | 3.91 | 3.14 | 0.94 |
| Γ- Γ **R** | 2.98 | 2.71 | 0.93 |
| Γ- Γ **S** | 2.94 | 2.61 | 0.95 |
| Γ- Γ **U** | 2.28 | 2.11 | 0.94 |
| Γ- Γ **Z** | 3.46 | 2.36 | 0.92 |

Further, we have calculated the values of α for the LFO compound using BoltzTraP code. The calculated α data in 300-1200 K temperature range is shown in Fig. 2 and found to be positive in the entire temperature range. For the comparison purpose the experimental data reported by Minh *et al* is also inserted in the Fig. 2. The experimental data is reported at 100 K temperature interval in the 573 to 1173 K temperature range. The experimental data have good agreement with the calculated values. Generally, for the insulating systems, the magnitude of the α gets decreases with temperature. The experimentally reported data have fluctuations in the α values at some temperature. These fluctuations can be due to the measurement conditions. Still, the net decreasing trend in the experimental value of α is consistent with the temperature dependent behavior of theoretically calculated values. The values of α in oxide system also depend on the synthesis condition of the material and it is also highly sensitive to the sample environment during the measurement condition. Therefore, more experimental measurements of α in the temperature range under study are highly desirable.

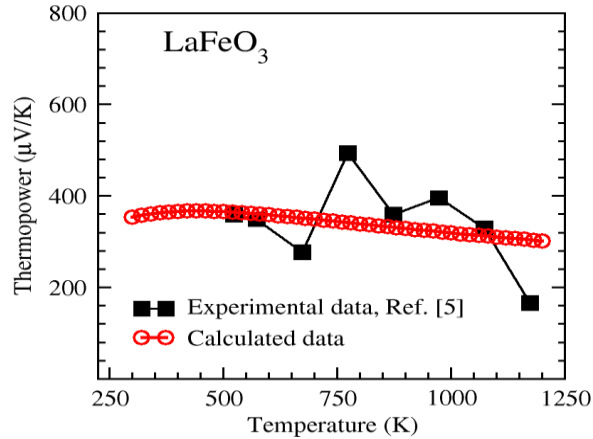

**FIGURE 2.** Temperature dependance of thermopower data for LaFeO$_3$ compound. Calculated values (red open circle with line) and experimental data (black solid square with line) from the Ref [5].

## CONCLUSION

In conclusion, we have investigated the electronic structure and thermopower behavior of LaFeO$_3$ compound using DFT+U method. The calculations on anti-ferromagnetic ground state structure of the system gives an insulating gap of ~2 eV, which is in accordance to the experimentally observed energy gap. From the dispersion curves, the calculated values of effective mass of holes in VB are found to be larger than the effective mass of electrons in CB. Especially, the effective mass of holes along Y, X, T and Z directions in the band *2* of VB are ~4 times than that of electrons effective mass in the band *1* of CB. The large value of effective mass of holes in doubly degenerate bands *2* and *3* in the VB are mainly responsible for showing the large and positive values of thermopower by this compound. The calculated values of thermopower using BoltzTraP code in the temperature range 300-1200 K are positive and found to be in good agreement with the experimental data reported in the literature. The present study shows that the theoretical tools are very effective in explaining the high temperature thermopower behavior of LaFeO$_3$ compound.